\documentclass[reprint, superscriptaddress, secnumarabic, amssymb, nobibnotes, aps, prl]{revtex4-1}

\usepackage{graphicx}
\usepackage{epstopdf}
\usepackage[T1]{fontenc}
\usepackage[latin9]{inputenc}
\usepackage{amsbsy}
\usepackage{gensymb}
\usepackage[T1]{fontenc}
\usepackage[latin9]{inputenc}
\usepackage{amsmath}
\usepackage{amssymb}
\usepackage{bbm}
\usepackage{physics}
\usepackage{braket}
\usepackage{xcolor}
\allowdisplaybreaks
\usepackage{graphicx}
\usepackage[colorlinks=true]{hyperref}  
\hypersetup{
    bookmarks=true,         % show bookmarks bar?
    unicode=false,          % non-Latin characters 
    pdftoolbar=true,        % show Acrobat
    pdfmenubar=true,        % show Acrobat 
    pdffitwindow=false,     % window fit to page when opened
    pdfstartview={FitH},    % fits the width of the page to the window
    pdftitle={Superconductivity in Bi based Bi$_2$PdPt},    % title
    pdfauthor={, , ,},     % author
    pdfsubject={},   % subject of the document
    pdfcreator={},   % creator of the document
    pdfproducer={},  % producer of the document
    pdfkeywords={} {} {}, % list of keywords
    pdfnewwindow=true,      % links in new window
    colorlinks=true,       % false: boxed links; true: colored links
    linkcolor=blue, %red,          % color of internal links (change box color with linkbordercolor)
    citecolor=blue,        % color of links to bibliography
    filecolor=magenta,      % color of file links
    urlcolor=blue           % color of external links
} 
\usepackage[normalem]{ulem}

\newcommand{\equref}[1]{Eq.~(\ref{#1})}

\newcommand{\figref}[1]{Fig.~\ref{#1}}

\begin{document}
\title{\textrm{Superconductivity in Bi based Bi$_2$PdPt}}
\author{A. Kataria}
\affiliation{Department of Physics, Indian Institute of Science Education and Research Bhopal, Bhopal, 462066, India}
\author{T. Agarwal}
\affiliation{Department of Physics, Indian Institute of Science Education and Research Bhopal, Bhopal, 462066, India}
\author{S. Sharma}
\affiliation{Department of Physics, Indian Institute of Science Education and Research Bhopal, Bhopal, 462066, India}
\author{D. Singh}
\affiliation{Department of Physics, Indian Institute of Science Education and Research Bhopal, Bhopal, 462066, India}
\author{S. Marik}
\affiliation{Department of Physics, Indian Institute of Science Education and Research Bhopal, Bhopal, 462066, India}
\author{R. P. Singh}
\email[]{rpsingh@iiserb.ac.in}
\affiliation{Department of Physics, Indian Institute of Science Education and Research Bhopal, Bhopal, 462066, India}
\date{\today}
\begin{abstract}
We report synthesis and properties of superconducting Bi$_2$PdPt, a new member of Bi-Pd based compounds known for their simultaneous existence of topological surfaces states and superconductivity. It crystallizes in a hexagonal structure having space group $P6_3/mmc$. A detailed investigation of the properties via transport, magnetization, and specific heat measurements confirm bulk superconductivity with transition temperature, \textit{T}$_{C}$ = 4.0(1) K in moderate coupling limit.
\end{abstract}
\maketitle

\section{Introduction}

The topological aspects of electronic states provide a thrust in the latest research of condensed matter physics. With the realization of spin polarization in these electronic states, various exotic quantum phases emerge, such as topological insulator, topological semimetals, and topological superconductor (TSC) \cite{Ts,ts1,ts2}. Particularly, topological superconductors attracted immense attention due to the potential appearance of exotic quasiparticles called Majorana fermions and their application in topological quantum computing. A topological superconductor is characterized by a bulk superconducting gap and Majorana zero modes at the symmetry protected topological surface states (TSSs) \cite{ts2}. The various proposed ways to realize TSC include carrier doped topological insulator and superconductor-topological insulator heterostructures \cite{MF,Mf1} which introduce the challenge to prepare high quality samples and observe interfacial phenomenons. Bulk compounds where non-trivial band structure and superconductivity intrinsically coexists emerge as a promising route to realize topological superconductivity. 

Bi-based materials are known for the presence of TSSs \cite{Bi-sb,Bi-se,TBS,TBS1,TBT,BPST,PBT,HH}. The  Bi-based binary systems $\alpha$-PdBi, $\alpha$-PdBi$_2$, and $\beta$-PdBi$_2$ are among the promising bulk topological superconductivity candidates. Notably, in $\alpha$-PdBi$_2$ and non-centrosymmetric $\alpha$-PdBi, Rashba surface states near the Fermi level, $E_{\mathrm{F}}$ were theoretically predicted and experimentally verified via angle-resolved photoemission spectroscopy (ARPES) with intrinsic superconductivity \cite{a-PB2,a-PB2_1,a-PB2_2,a-PB2_3,BP1,BP2}. Moreover, the observed non-helical spin texture near $E_{\mathrm{F}}$ for $\beta$-PdBi$_2$ combined with the various reports of fully gapped multiband superconductivity in bulk, suggests the realization of Majorana fermions at the surface states \cite{b-PB2_A,b_PB2,b-PB2_s,Msc,Msc1}.

Recently, PtBi$_2$ displays a large anisotropy for the in-plane and inter-plane transport and is stated as a bulk topological metal with pressure-induced superconductivity at 2.0 K \cite{Ptbi,Ptbi2}. It indicates that the vast family of Bi compounds is associated with various aspects of topological electronic states and accompanied by superconductivity, which demands detailed investigation in this family. Further, the spin-orbit coupling (SOC) strength affect the Fermi surface to capture the non-trivial band topology \cite{Bi-se,TI} and also impact the low-energy properties like superconductivity \cite{socsc}. Thus, the superconductivity with high SOC strength and the possible non-trivial band structure may direct unconventionality in the superconducting ground state. As SOC is directly proportional to the fourth power of atomic number $Z$, introducing a heavier element will increase the SOC strength of the system.

We have synthesized a new member of the Bi compound family, Bi$_2$PdPt, where a Pt atom partially replaces a Pd atom from PdBi. Powder XRD and Laue diffraction were performed to investigate the crystal structure and orientation. The superconducting and normal state properties of Bi$_2$PdPt studied using transport, magnetization, and specific heat measurements. Type-II superconductivity in moderate coupling limit is established in Bi$_2$PdPt.

\begin{figure*}[ht!] %{r}{0.5\textwidth}
\includegraphics[width=2.0\columnwidth, origin=b]{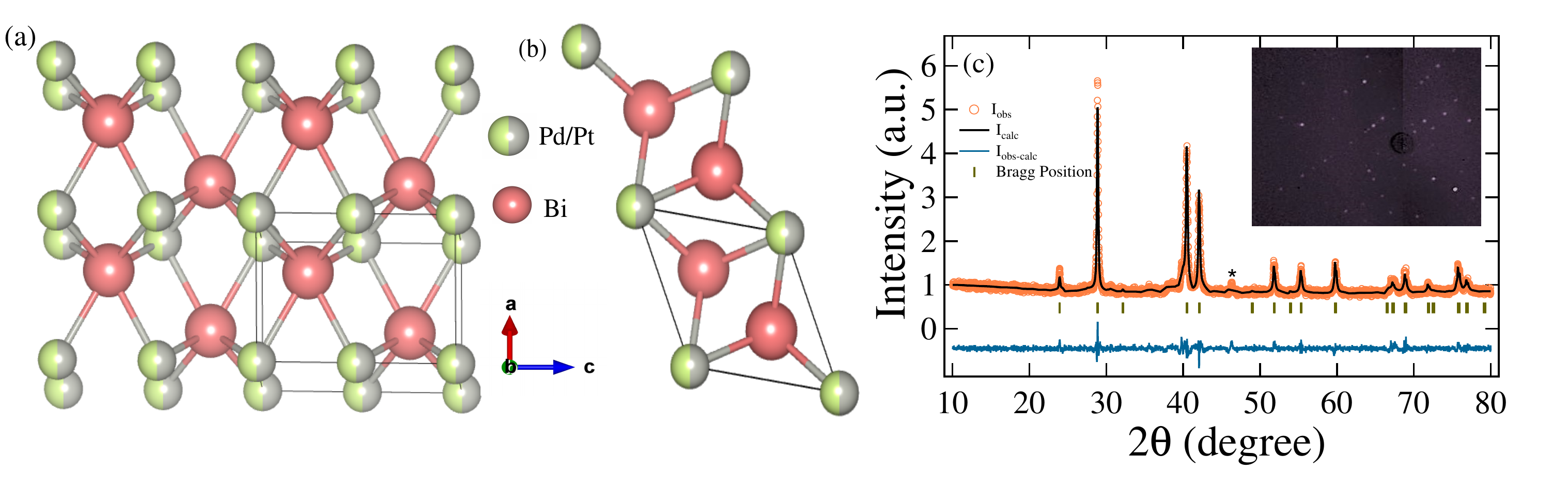}
\caption{\label{Fig1:XRD} The NiAs hexagonal crystal structure of Bi$_2$PdPt is shown with solid orange ball as Bi atoms and half green or grey solid ball as Pd/Pt atoms. (a) Shows the side view of the crystal structure and (b) shows the top view. (c) Rietveld refined pattern, represented by black line, of XRD recorded at room temperature (orange open circle) for Bi$_2$PdPt is shown. Inset shows the Laue diffraction pattern of the single-crystal. }
\end{figure*}

\section{Experimental Details}

Single-crystal of Bi$_2$PdPt was prepared by the modified Bridgman method. Firstly, the polycrystalline sample was prepared using the solid-state reaction method by taking a stoichiometric mixture of required elemental powder Bi (99.99\%), Pd (99.95\%) and Pt (99.9\%). The palletized form of the mixture was subsequently heated at 700$\degree$C for four days in a vacuum sealed tube, which was then transferred in a conical end quartz ampoule for single-crystal growth. The ampoule dwelled at 700$\degree$C for 12 h and slowly cooled up to 650$\degree$C at the rate of 0.4$\degree$C/hr, followed by the water quenching. The obtained single-crystal is extremely brittle with a metallic lustre. Mechanically cleaved single-crystals were used for all the measurements. Powder x-ray diffraction data for grinded single-crystal was collected at room temperature using PANalytical X'Pert powder diffractometer  with Cu$K_{\alpha}$ ($\lambda$ = 1.5406 \AA) radiation. The Laue diffraction pattern was recorded using the Photonic-Science Laue camera system for single-crystal orientation and quality determination. The magnetic measurements were performed on the vibrating sample magnetometer (VSM) option of the Quantum Design Magnetic Property Measurement System (MPMS-3). AC transport and specific heat measurements with two-$\tau$ relaxation method were performed using Quantum Design Physical Property Measurement System (PPMS) 9 T with an applied magnetic field parallel to the crystallographic [001] axis.

\section{Results and Discussion}

\subsection{Sample characterization}

Rietveld refinement of powder x-ray diffraction pattern of Bi$_2$PdPt was performed using FullProf software (shown in \figref{Fig1:XRD}) \cite{fp}. A small elemental impurity peak of Pt is observed shown by an asterisk in the \figref{Fig1:XRD}. The refinement confirms the crystallization of Bi$_2$PdPt in NiAs hexagonal crystal structure having space group $P6_3/mmc$, which is different from the structures reported for other superconducting compounds of the family, $\alpha$-BiPd (non-centrosymmetric) and $\alpha$-, $\beta$-PdBi$_2$, PtBi$_2$ (layered structure) \cite{a-PB2,b_PB2,BP1,Ptbi}. 

The structure of Bi$_2$PdPt consists of Bi atoms that lie inside the hexagonal geometry while Pd and Pt atoms share the edges sites with equal probability, as shown in the \figref{Fig1:XRD}(a) and (b). Detailed refinement results with the associated Wyckoff position of each atom are summarized in Table I. The Laue diffraction pattern of different crystals of the same batch depicts the orientation of the crystal along [001] direction, and bright spots tell the quality of the single-crystal as shown in the inset of \figref{Fig1:XRD}(c).

\begin{table}[h!]
\caption{Structure parameters of Bi$_2$PdPt obtained from the Rietveld refinement of XRD}
\begin{tabular}{l r} \hline\hline
Structure& Hexagonal\\
Space group&        $P6_3/mmc$\\ [1ex]
Lattice parameters\\ [0.5ex]
a = b(\text{\AA})&  4.294(5)\\
   c (\text{\AA}) & 5.567(1)
\end{tabular}
\\[1ex]

\begingroup
\setlength{\tabcolsep}{4pt}
\begin{tabular}[b]{c c c c c c}
Atom&  Wyckoff position& x& y& z& Occupancy\\[1ex]
\hline
Bi& 2c& 0.333& 0.666& 0.250& 0.0839\\             
Pd& 2a& 0& 0& 0& 0.0417\\                       
Pt& 2a& 0& 0& 0& 0.0416\\
[1ex]
\hline
\end{tabular}
\par\medskip\footnotesize
\endgroup
\end{table}

\subsection{Superconducting and normal state properties}

\begin{figure} %{r}{0.5\textwidth}
\includegraphics[width=1.0\columnwidth, origin=b]{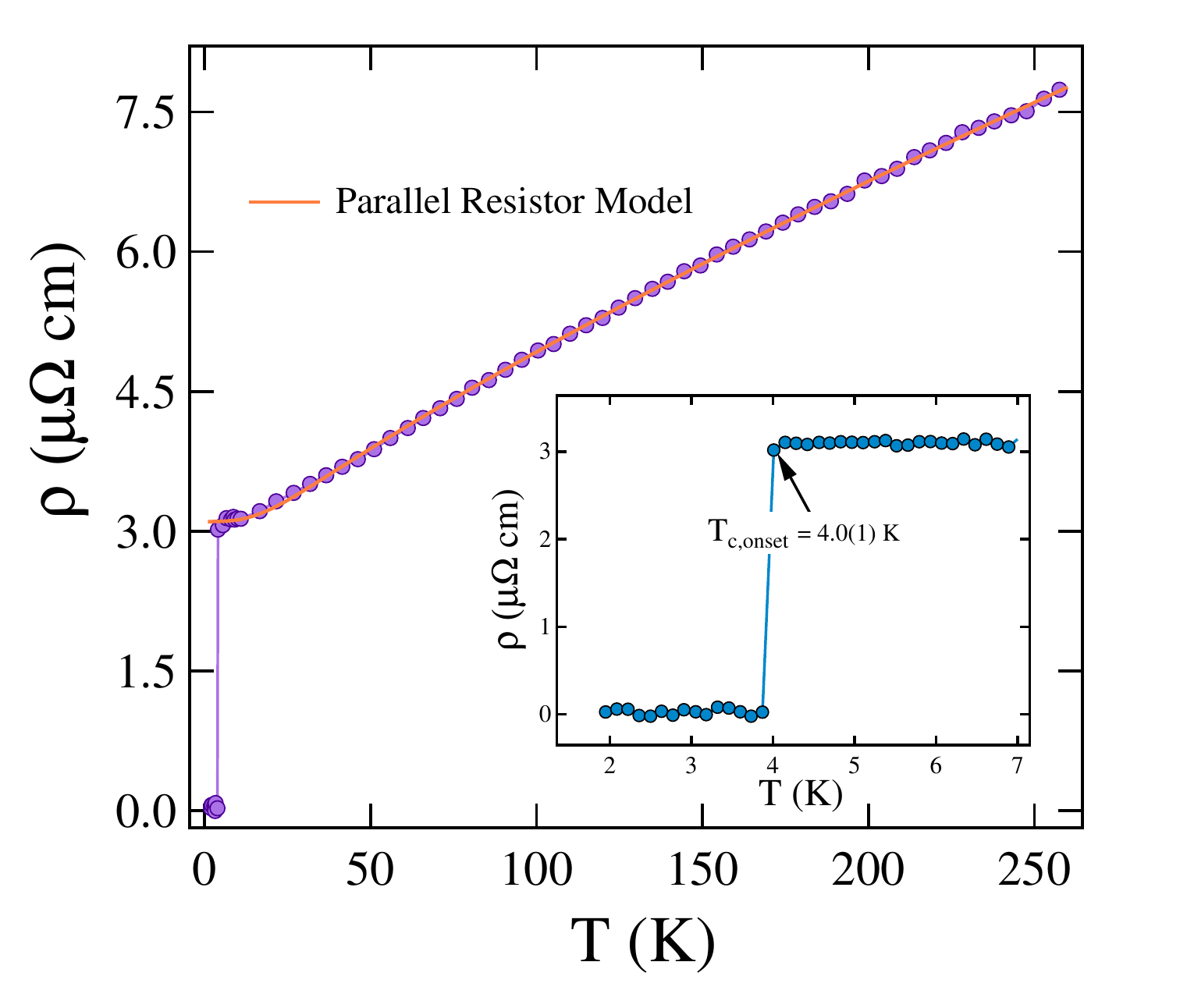}
\caption{\label{Fig3:Res} Temperature dependence of resistivity in zero field fitted using the parallel resistor model above transition temperature is shown in main panel. The inset shows the limited temperature region showing superconducting transition at $T_{C,onset}$ = 4.0(1) K. }
\end{figure}

In-plane AC resistivity at zero magnetic field in temperature range, 1.8 K to 250 K, was measured for single-crystal of Bi$_2$PdPt, as shown in \figref{Fig3:Res}. A sharp resistivity drop from the normal state resistivity at $T_{C,onset}$ = 4.0(1) K with the width of 0.1 K is registered as a superconducting transition. Above transition temperature resistivity behaviour of Bi$_2$PdPt shows slight deviation from linear behaviour, which is described by Wilson's theory \cite{WT}. However, the high value of absolute resistivity with a saturated trend is allied with $\alpha$-BiPd and other A-15 compounds where a parallel resistor model is implanted \cite{SCbp}. The resistivity behaviour for Bi$_2$PdPt is characterised using parallel resistor model described by Wisemann as \cite{Wm},

\begin{equation}
\frac{1}{\rho(T)} =\frac{1}{\rho_s} + \frac{1}{\rho_i(T)}    \label{eqn1:rho}
\end{equation}

where $\rho_s$ is saturated resistivity independent of the temperature and $\rho_i(T)$ is the ideal contribution consisting two terms which expressed as,  

\begin{equation}
\rho_i(T) = \rho_{i,0} + C\left(\frac{T}{\Theta_R}\right)^n \int_{0}^{\Theta_R/T}\frac{x^n}{(e^x-1)(1-e^{-x})}dx\\
\label{eqn2:rbg}
\end{equation}
\\

here the first term $\rho_{i,0}$ is temperature-independent residual resistivity, and the second term represents the temperature dependent contribution including the effect of electron-phonon scattering stated by the generalized Bloch-Gr\"uneisen expression \cite{bg,WT}. $\Theta_R$ is the Debye temperature from resistivity measurement, and $C$ is a material-dependent quantity. Best fitting of Bi$_2$PdPt for $n$ = 3 yields Debye temperature, $\Theta_R$ = 123(3) K, $C$ = 5.70(2) $\mu\ohm$-cm, $\rho_s$ = 47(3) $\mu\ohm$-cm and $\rho_{i,0}$ = 3.32(1) $\mu\ohm$-cm. The estimated residual resistivity ratio, $\rho(250K)/\rho(5K)$ = 2.5 is much smaller than the reported values for the other Bi-Pd compounds, which can be possibly due to the increase in disorder of the system by Pt presence \cite{SCbp,scpb2}. 

\begin{figure} [b] %{r}{0.5\textwidth}
\includegraphics[width=1.0\columnwidth, origin=b]{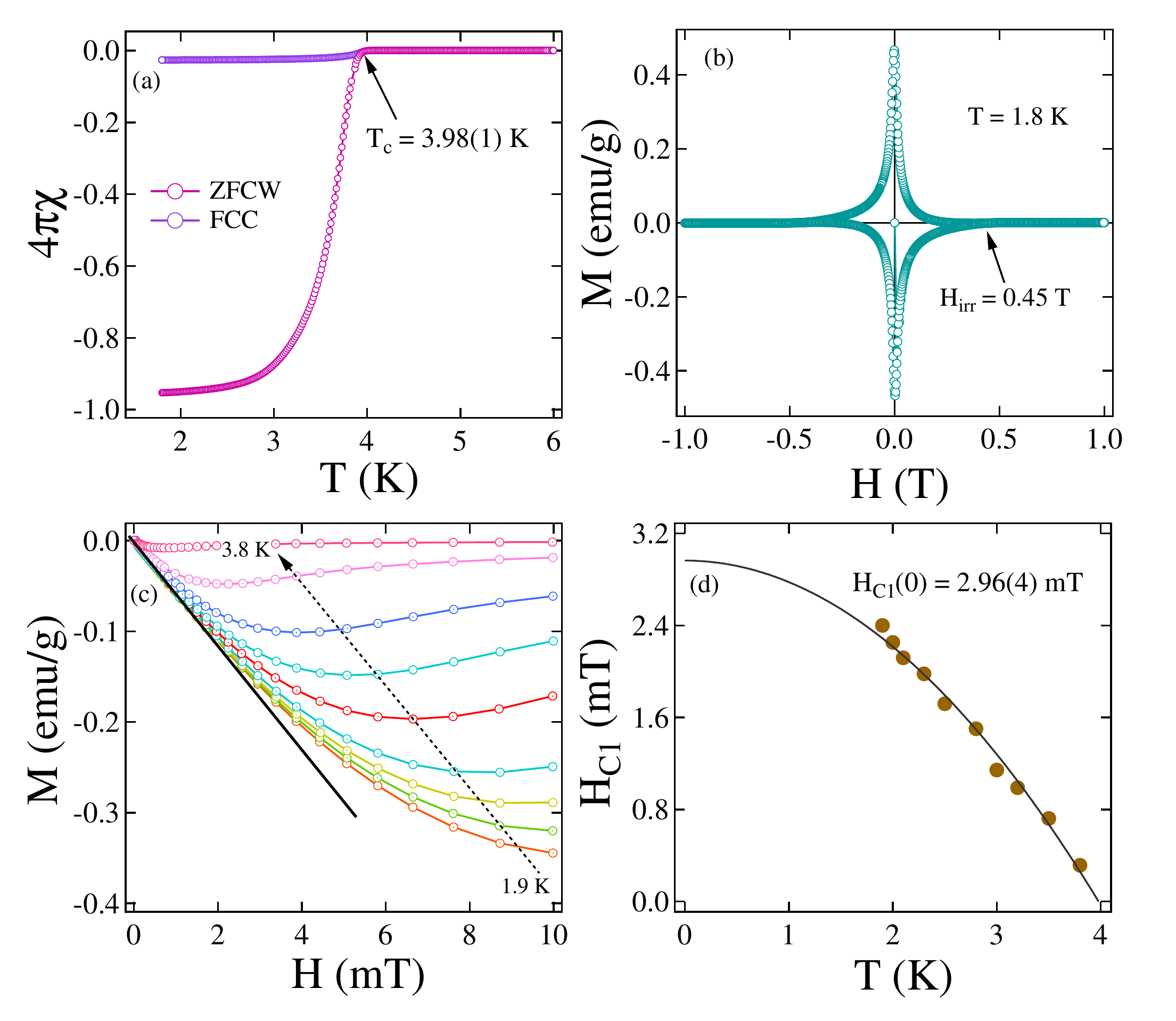}
\caption{\label{Fig4:MT} a) Shows the superconducting transition in 1 mT magnetic field at $T_{c,onset}$ = 3.98(1) K. (b) Magnetization hysteresis loop at 1.8 K. (c) Magnetic moment variation under magnetic field at different temperature upto transition temperature. (d) Lower critical field with temperature fitting using GL equation \equref{eqn2:Hc2}.} 
\end{figure}

\hfill \newline
Magnetization versus temperature measurement at 1 mT magnetic field in zero-field-cooled warming (ZFCW) and field-cooled cooling (FCC) mode is shown in \figref{Fig4:MT}(a). Curve shows a diamagnetic onset at temperature $T_{C}$ = 3.98(1) K perpendicular to crystallographic [001] axis of single-crystalline Bi$_2$PdPt. The poor overlapping of the ZFCW and FCC curve depicts the presence of a strong pinning centre in the sample with superconducting volume fraction close to 96\% certain the bulk property. Magnetization hysteresis loop under an applied magnetic field of $\pm$ 1 T at temperature, 1.8 K is shown in \figref{Fig4:MT}(b). The loop represents the conventional type-II behaviour along with an irreversible nature of magnetization below $H_{irr}$ = 0.45 T, above which unpinning of the vortices start taking place.

\begin{figure}%{r}{0.5\textwidth}
\includegraphics[width=1.0\columnwidth, origin=b]{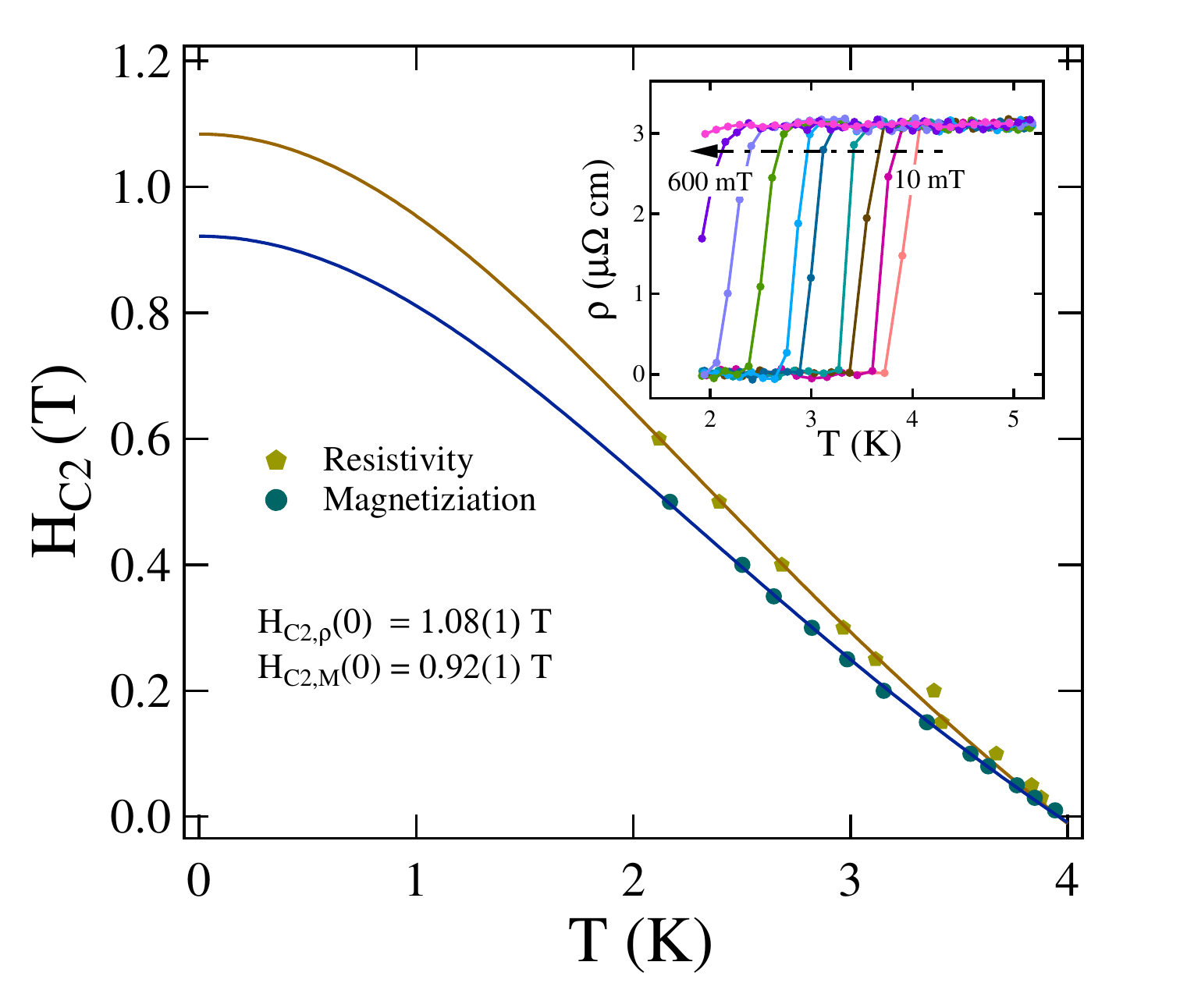}
\caption{\label{Fig5:hc2} The upper critical field variation with temperature is determined via magnetization and resistivity measurements and fitted by using the equation \equref{eqn2:Hc2} is shown. Inset shows the resistivity drop temperature variation with applied magnetic field.} 
\end{figure}

Further, to estimate the lower critical field value $H_{C1}$, the magnetization variation in the applied magnetic field at different isotherms up to transition temperature, $T_C$, were analyzed. $H_{C1}$ for each isotherm is considered as a magnetic field value where the curve starts to deviate from the linear Meissner state, as shown in \figref{Fig4:MT}(c). The temperature evolution of the lower critical field is fitted with the Ginzburg-Landau (GL) equation,
\\
\begin{equation}
H_{C1}(T)=H_{C1}(0)\left[1-\left(\frac{T}{T_{C}}\right)^{2}\right]
\label{eqn1:Hc1}
\end{equation} 

which gives $H_{C1}(0)$ = 2.96(4) mT for $T_C$ = 3.98(1) K (\figref{Fig4:MT}(d)). The upper critical field $H_{C2}$ is evaluated by the observed shifts in superconducting transition temperature with the increasing applied magnetic field. The shift in the transition temperature, $T_C$ is marked from magnetization at onset and for resistivity measurements at 90$\%$ drop from normal state resistivity value ($T_{C,90\%}$), as shown in the inset of \figref{Fig5:hc2}. The upper critical field $H_{C2}(0)$ is estimated from the expression, 
\\
\begin{equation}
H_{C2}(T) = H_{C2}(0)\left[\frac{(1-t^{2})}{(1+t^2)}\right]. 
\label{eqn2:Hc2}
\end{equation}

here $t = T/T_C$, reduced temperature. The fitted data yields $H_{C2,M}(0)$ = 0.92(1) T and $H_{C2,\rho}(0)$ = 1.08(1) T from magnetization, and resistivity measurements, respectively displayed in \figref{Fig5:hc2}.

\begin{figure} %{r}{0.5\textwidth}
\includegraphics[width=1.0\columnwidth, origin=b]{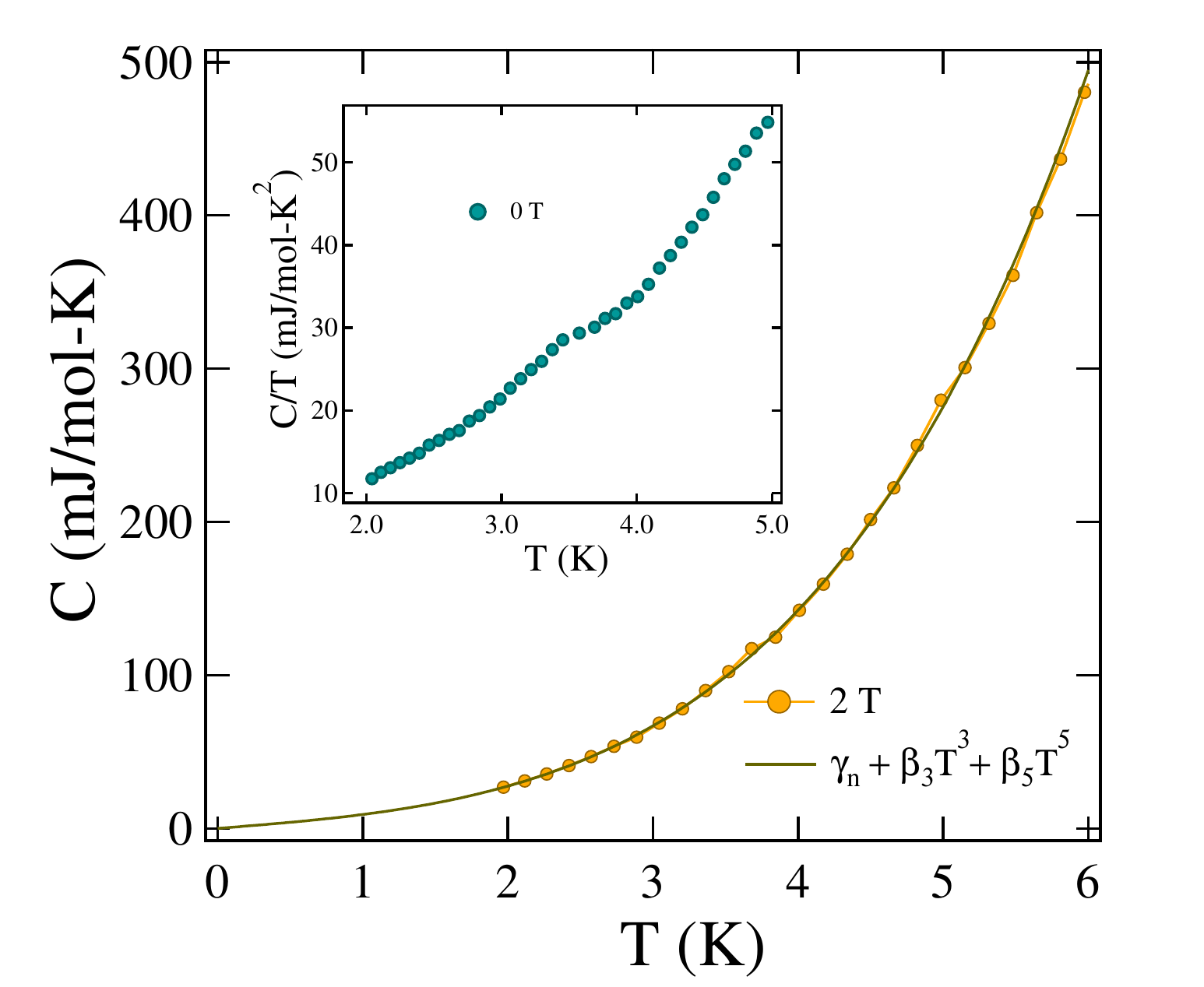}
\caption{\label{Fig6:SH} The temperature dependence specific heat of the Bi${_2}$PdPt represented in $C$ vs $T$ form at 2 T magnetic field is fitted with the equation \equref{eqn2:SH1}. Inset shows the $C$ vs $T$ curve at zero magnetic field curve where anomaly is clearly observed.}
\end{figure}

\hfill\newline
In addition to the resistivity and magnetization measurements, specific heat is also measured at zero field and 2 T (well above the upper critical field) under temperature variation, shown in \figref{Fig6:SH}. A small anomaly in 0 T specific heat measurement is inscribed around 3.93(7) K, as shown in the inset of \figref{Fig6:SH}, which is in allied with the superconducting transition temperature recorded from resistivity and magnetization measurements. Whereas the measurement recorded at 2 T presents the normal state of Bi$_2$PdPt. The normal state specific data is fitted using the Debye relation,

\begin{equation}  
C = \gamma_{n} T+\beta_{3} T^{3} + \beta_{5}T^{5}
\label{eqn2:SH1}    
\end{equation} 

where $\gamma_n$ is Sommerfeld coefficient corresponding to electronic contribution, $\beta_3$ Debye constant represents phononic contribution, and $\beta_5$ produces the anharmonic contribution to the specific heat. The fitted data produces $\gamma_n$ = 5.1(6) mJmol$^{-1}$K$^{-2}$, $\beta_3$ = 1.72(6) mJmol$^{-1}$K$^{-4}$ and $ \beta_{5} $ = 11.2(1) $\mu$Jmol$^{-1}$K$^{-6}$ (\figref{Fig6:SH}). The Sommerfeld coefficient value matches well with the $\alpha$-PdBi reported value, but a large deviation from $\alpha$- and $\beta$-PdBi$_2$ is observed \cite{SCbp,scpb2,p-pb2_1}. According to the free-electron theory, the Sommerfeld coefficient is used to evaluate the density of state at Fermi level as per the relation  $\gamma_{n}=\left(\frac{\pi^{2}k_{B}^{2}}{3}\right)D_{C}(E_{\mathrm{F}}) $, where $k_B$ = 1.38$\times$10$^{-23}$ J K$^{-1}$ is Boltzmann constant. The $D_C(E_{\mathrm{F}})$ is calculated to be 2.1(2) states eV$^{-1}$ f.u.$^{-1}$. Further, considering the Debye model, the Debye temperature $\Theta_D$ is estimated to be 165(2) K for $\beta_3$ = 1.72(6) mJ mol$^{-1}$ K$^{-4}$ using the expression $\theta_{D}= \left(\frac{12\pi^{4}RN}{5\beta_{3}}\right)^{\frac{1}{3}}$, where $R$ = 8.314 J mol$ ^{-1} $K$ ^{-1} $ is a gas constant. The calculated Debye temperature value is consistent with the value obtained from parallel resistor model. The average electron-phonon coupling constant, $\lambda_{e-ph} $ is calculated by implanting McMillan's theory given as \cite{mcm},

\begin{equation}
\lambda_{e-ph} = \frac{1.04+\mu^{*}\mathrm{ln}(\theta_{D}/1.45T_{C})}{(1-0.62\mu^{*})\mathrm{ln}(\theta_{D}/1.45T_{C})-1.04 }
\label{eqn6:Lambda}
\end{equation}

where $\mu^*$ varies in the range of 0.1 $\leq \mu^* \leq $ 0.15, a material-specific value accounts for screened Coulomb repulsion. Considering $T_C$ = 3.93(7) K, $\theta_D$ = 165(2) K and $\mu^*$ = 0.1 (as speculate for Bi-Pd binary compounds) obtained $\lambda_{e-ph}$ = 0.64(1), which puts Bi$_2$PdPt in the moderately coupled superconductor regime respective to other Bi-Pd compounds $\alpha$-, $\beta$-PdBi$_{2}$, and $\alpha$-PdBi \cite{SCbp,p-pb2_1}. 

\subsection{Superconducting parameters and Uemura Plot}

The external applied magnetic field puts a limiting field on the superconductor, portrayed independently by spin paramagnetic and orbital limiting effects. According to the BCS theory, spin paramagnetic effect comes from pairing breaking due to Zeeman effect defined as the Pauli limiting field, $ H_{C2}^{P}(0)$ = 1.86 $T_{C} $ \cite{Pauli_1,Pauli_2}. For Bi$_2$PdPt with $T_C$ = 3.98(1) K, yields $ H_{C2}^{P}(0)$ = 7.40(2) T. The orbital limit for an upper critical field refers to the critical field where vortex cores begin to overlap. The orbital limiting field is given by Werthamer-Helfand-Hohenberg (WHH) expression \cite{WHH_1,WHH_2}, 

\begin{equation}
H_{C2}^{orbital}(0) = -\alpha T_{C}\left.\frac{dH_{C2}(T)}{dT}\right|_{T=T_{C}}
\label{eqn3:HHH}
\end{equation}
\\
where $\alpha$ = 0.69 for dirty limit and 0.73 for clean limit superconductors. The initial slope $\frac{-dH_{C2}(T)}{dT} $ at $T = T_{C}$ is estimated to be 0.23(1) T/K which gives $ H_{C2}^{orbital}(0)$ = 0.64(1) T considering the dirty limit. Moreover, the relative strength of Pauli and orbital limiting field is measured by Maki parameter defined by the relation $\alpha_{M} = \sqrt{2} H_{C2}^{orb}(0)/H_{C2}^{P}(0)$, evaluates $\alpha_{M}$ = 0.12(2) \cite{maki}. The small value of Maki parameter indicates the negligible effect of Pauli limiting field for Bi$_2$PdPt. 

Further, the fundamental superconducting length parameters is calculated from upper, $H_{C2}(0)$, and lower, $H_{C1}(0)$ critical field values. The Ginzburg-Landau coherence length ($\xi_{GL}(0)$) is estimated from $\xi_{GL}(0) = \sqrt{\frac{\Phi_0}{2\pi H_{C2}(0)}}$ \cite{Coh_Leng}, here $\Phi_{0}$  = 2.07 $\times$ 10$^{-15}$ Tm$^{2}$ is the magnetic flux quantum. Using $H_{C2,M}(0) $ = 0.92(1) T, $\xi_{GL}(0)$ is evaluated to be 190(1) \text{\AA}. The equation \cite{MIB},

\begin{equation}
H_{C1}(0) = \frac{\Phi_{0}}{4\pi\lambda_{GL}^2(0)}\left(\mathrm{ln}\frac{\lambda_{GL}(0)}{\xi_{GL}(0)}+0.12\right)   
\label{eqn4:PD}
\end{equation} 

provides the Ginzburg-Landau penetration depth, $\lambda_{GL}(0)$ = 4231(76) \text{\AA} for $H_{C1}(0)$ = 2.96(4) mT  and $\xi_{GL}$(0) = 190(1) \text{\AA}. Moreover, the ratio $\frac{\lambda_{GL}(0)}{\xi_{GL}(0)}$ gives the Ginzburg-Landau parameter, $\kappa_{GL}$ = 22.2(5) which is much greater than $1/\sqrt{2}$ indicating a strong type-II superconductivity in Bi$_2$PdPt. The thermodynamic critical field $H_{C}$, is evaluated by using the expression \cite{MIB}; $H_{C1}(0)H_{C2}(0) = H_{C}^2\mathrm{ln}\kappa_{GL}$, yielding $H_{C}$ = 29(1) mT.

For Bi$_2$PdPt, in order to determine the rest of the superconducting parameters including London penetration depth $\lambda_L$, electronic mean free path $l_e$ and to verify dirty limit superconductivity, a set of equation is used assuming the spherical Fermi surface. The electronic mean free path, $l_e$ is estimated from the quasiparticle number density, $n$ and Sommerfeld coefficient, $\gamma_n$. The quasiparticle number density of Bi$_2$PdPt has been extracted from Hall measurement, $n = 2.45(2) \times 10^{28}$ m$^{-3}$ (more details in appendix). The Fermi vector, $k_{\mathrm{F}}$ is evaluated using $k_{\mathrm{F}}$ = (3$\pi^{2}n)^{1/3}$, which gives $k_{\mathrm{F}}$ = 0.90(1) \AA$^{-1}$. The effective mass, $m^*$ is calculated from Sommerfeld coefficient, $m^* = {(\hbar k_{\mathrm{F}})^{2}\gamma_{n}/\pi^{2}n k_B^{2}}$. For $k_B$ is Boltzmann constant and $\gamma_{n}$ = 95.51 J m$^{-3}$ K$^{-2}$, $m^*$ = 2.0(3) $m_e$.  According to Drude Model \cite{Coh_Leng}, Fermi velocity is determined from the relation $v_{\mathrm{F}} = \hbar k_{\mathrm{F}}/m^*$ = 5.1(7) $\times$ 10$^5$ m s$^{-1}$ and scattering time from $\tau^{-1} = ne^2\rho_{0}/m^{*}$, which further used to calculate electronic mean free path, $l_e = v_{\mathrm{F}} \tau$ providing $l_e$ = 455(68) \text{\AA}. 

In consideration of BCS theory, the coherence length  $\xi_{0}$ is defined as $0.18\hbar v_{\mathrm{F}}/k_BT_c$ providing $\xi_0$ = 1756(147) \text{\AA}. The ratio of $\xi_0/l_e$ = 3.9 is greater than 1, implying Bi$_2$PdPt is a dirty limit superconductor. Further in the dirty limit, the London penetration depth $\lambda_{L}$ \cite{Umera_Ref}, is written as,

\begin{equation}
\lambda_{L}=\left(\frac{m^{*}}{\mu_{0}n e^{2}}\right)^{1/2}.
\label{eq9:l}\\
\end{equation}

From the above expression the estimated $\lambda_{L}$ is 483(36) \text{\AA}. The order of calculated parameters matches well with $\beta$-PdBi$_2$ where dirty limit superconductivity is observed \cite{Msc} whereas $\alpha$-PdBi lies in a clean type-II superconducting limit \cite{SCbp}.

\begin{figure}
\includegraphics[width=1.0\columnwidth]{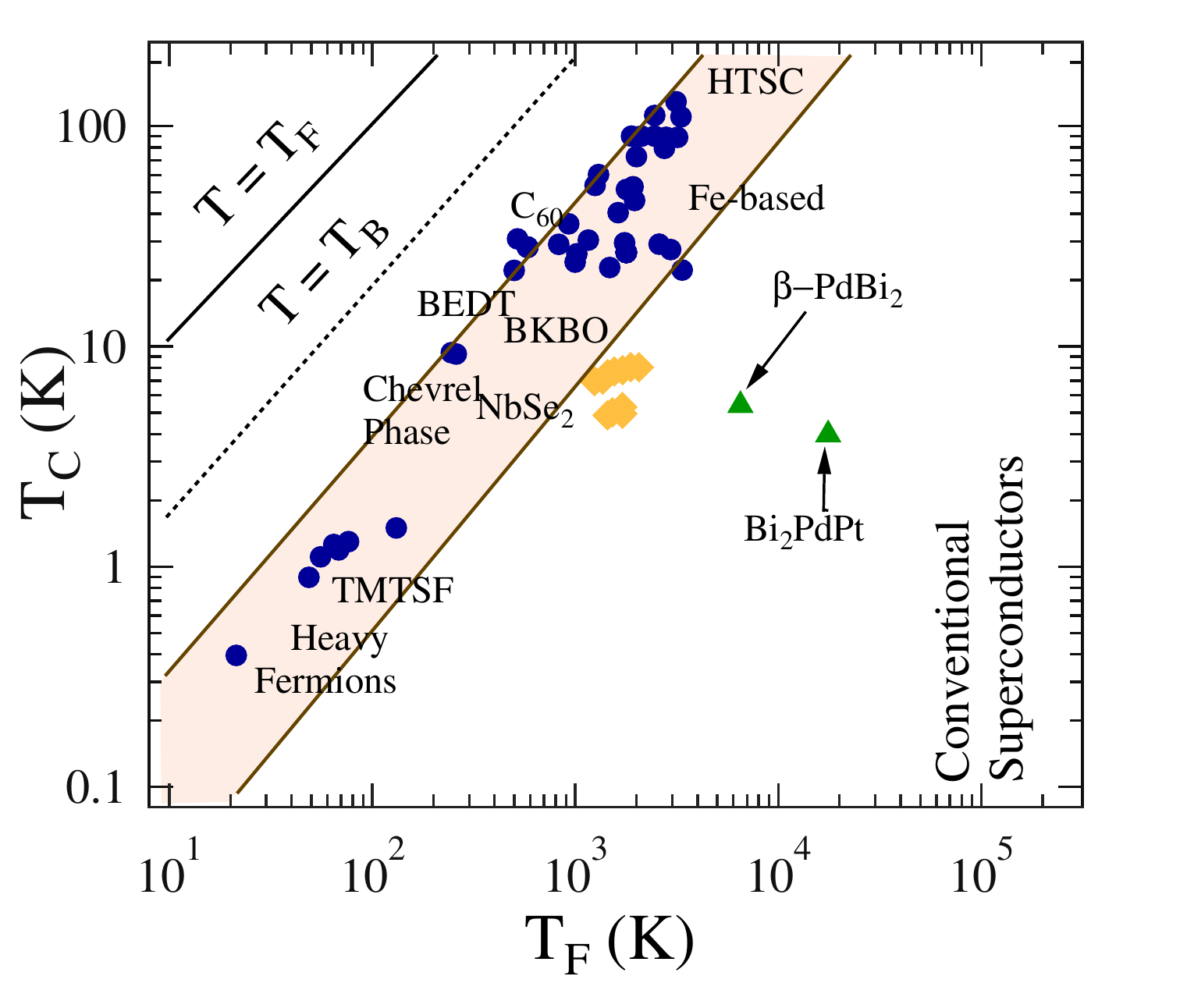}
\caption{\label{Fig8:UP} The Uemura plot showing the superconducting transition temperature $T_C$ vs the effective Fermi temperature $T_{{\mathrm{F}}}$ for Bi$_2$PdPt with $\beta$-PdBi$_2$, shown as solid green circle. Other data points represent the different families of superconductors together with the shaded region representing the families of unconventional superconductors. The dashed line corresponds to the Bose-Einstein condensation temperature $T_B$.}
\end{figure}

Moreover, to classify the Bi$_2$PdPt in the context of other superconductors, Uemura plot \cite{Umera} of $T_C$ with effective Fermi temperature, $T_{\mathrm{F}}$ has been plotted, which is used to define the character of unconventionality of a superconductor based on the ratio of $\frac{T_{C}}{T_{\mathrm{F}}}$. For 3D spherical Fermi surface, Fermi temperature is calculated using \cite{Tf},

\begin{equation}
k_{B}T_{{\mathrm{F}}} = \frac{\hbar^{2}}{2}(3\pi^{2})^{2/3}\frac{n^{2/3}}{m^{*}}, 
\label{eqn13:tf}
\end{equation}

\begin{table}[h!]
\caption{Parameters in the superconducting and normal state of Bi$_2$PdPt and $\alpha$-BiPd}
\begingroup
\setlength{\tabcolsep}{8
pt}
\begin{tabular}{c c c c} 
\hline\hline
Parameters & Unit & Bi$_2$PdPt  & $\alpha$-BiPd \cite{SCbp}\\ [1ex]
\hline
$T_{C}$& K& 4.0 & 3.8\\             
$H_{C1}(0)$& mT& 2.96 & 12.0\\                       
$H_{C2,M}(0)$& T& 0.92 & 0.8\\
$H_{C2}^{P}(0)$& T& 7.40 & 7.07\\
$H_{C2}^{Orb}(0)$& T& 0.64 \\
$\xi_{GL}(0)$& \text{\AA}& 190 & 170\\
$\lambda_{GL}(0)$& \text{\AA}& 4231 & 1792\\
$k_{GL}$& &22.2 & 7.6\\
$\gamma_{n}$&  mJ mol$^{-1}$ K$^{-2}$& 5.1 & 4.0 \\   
$\theta_{D}$& K& 165.0 & 169.0\\
$\xi_{0}/l_{e}$&   &3.9 & 0.01\\
$v_{\mathrm{F}}$& 10$^{5}$ m s$^{-1}$& 5.1 &\\
$n$& 10$^{28}$m$^{-3}$& 2.45&\\
$T_{\mathrm{F}}$& K& 17576&\\
$m^{*}$/$m_{e}$&  & 2.0\\
[1ex]
\hline\hline
\end{tabular}
\endgroup
\end{table}

where $n$ is the quasiparticle number density per unit volume, and $m^{*}$ is the effective mass of quasiparticles. The $T_F$ value is estimated to be $T_{\mathrm{F}}$ = 17576(104) K, by considering the respective values of parameters from Table II. The ratio $\frac{T_{C}}{T_{\mathrm{F}}}$ = 0.0002 places this material, Bi$_2$PdPt close to other family member, $\beta$-PdBi$_2$ and outside the band of unconventional superconductor which includes iron based superconductors, cuprates and Chevrel phase \cite{Unconv_1,Unconv_2}. The conventional superconductors are considered to be on the right-hand side of the Uemura plot. Bi$_2$PdPt with the other member of the Bi-Pd compound family, $\beta$-PdBi$_2$ \cite{Msc} are shown by the solid green symbol in \figref{Fig8:UP}.

Among the Bi-Pd compound family, Bi$_2$PdPt is the highest possible SOC strength candidate, based on the presence of heavy constituent elements, establishes a strong foundation to exhibit intriguing properties. The different crystal structure NiAs type of Bi$_2$PdPt eliminates all the possibilities of any secondary superconducting phase presence of Bi-Pd/Pt compound, implementing the intrinsic bulk superconductivity. Hall measurement (see appendix) suggests dominated hole charge carries (10$^{28}$m$^{-3}$) same as for other Bi-Pd compounds, which indicate a very low effect in the electronic state with the Pt presence. The specific heat of Bi$_2$PdPt gives Sommerfeld coefficient 5.1(6) mJmol$^{-1}$K$^{-2}$, which is slightly greater than what is reported for $\alpha$-BiPd, suggests only a minimal increase in density of states at $E_{\mathrm{F}}$. However, the observed value is comparable to that of $\beta$-PdBi$_2$ and Pt$_{1.26}$Bi$_2$ \cite{pb}. Concluding neither substitution at Pd site nor increasing the Bi concentration predominately affects the density of states in Bi-Pd/Pt compounds. Despite the presence of heavy 5d Pt element, no change in phonon frequency, extracted from Debye temperature, is noted between Bi$_2$PdPt and $\alpha$-BiPd \cite{SCbp}.

Moreover, the bulk measurement of Bi$_2$PdPt is not sufficient to establish the gap symmetry of the sample. The possible explanation of low specific jump value can be due to multiband superconductivity with a single superconducting gap \cite{bi2pd}. The layered transition metal dichalcogenide superconductors have been well known to exhibit no anomaly in specific heat measurement \cite{tmc}. In addition, the low value of specific heat jump or no jump at all is also observed in many Bi-based half Heusler alloys (RPdBi), where superconductivity is attributed to the surface states \cite{BiHH}. Further, the similar behaviour is determined for the case of strongly or locally disordered systems such as quasi-skutterudites \cite{sku}. 

No deviation or anisotropy has been observed in the upper critical field of Bi$_2$PdPt, directing towards the bulk nature of the superconducting state and absence of multigap superconductivity. Apart from this, the variation of superconducting volume fraction with the different cooling procedures of sample preparation showed the grain boundaries effect in the sample (details in appendix). To fully understand the effect of increased SOC strength, the possible reason for the low value of specific heat jump and address any non-trivial band structure in the normal state of Bi$_2$PdPt detailed theoretical and experimental analysis of the superconducting ground state with band topology is required.

\section{CONCLUSION}

In summary, we synthesized the single-crystalline Bi$_2$PdPt and transport, magnetization, and specific heat measurements performed for detailed analysis of superconducting properties. The superconducting transition of Bi$_2$PdPt has been reported at 4.0(1) K with upper and lower critical field value, $H_{C1}$ = 2.96(4) mT and $H_{C2,M}$ = 0.92(1) T, fitted using Ginzburg-Landau equation. An anomaly was observed around transition temperature in specific heat measurement confirms the bulk superconductivity. The comparison of various normal and superconducting state parameters of Bi$_2$PdPt and $\alpha$-BiPd have been summarized in Table II. The Bi$_2$PdPt is classified as a moderately coupled type-II dirty limit superconductor. The estimated value of Fermi temperature, $T_{\mathrm{F}}$ places Bi$_2$PdPt close to the other member of the family, $\beta$-PdBi$_2$. Further microscopic measurements such as muon spectroscopy, scanning tunnelling microscope (STM), and angle-resolved photoemission spectroscopy (ARPES) are required to understand the superconducting ground state and the possible topological nature of Bi$_2$PdPt.

\section{Acknowledgments} A. Kataria acknowledges the funding agency Council of Scientific and Industrial Research (CSIR), Government of India, for providing SRF fellowship (Award No: 09/1020(0172)/2019-EMR-I). R.~P.~S.\ acknowledges Science and Engineering Research Board, Government of India for the Core Research Grant CRG/2019/001028.

\section{Appendix}

\subsection{Characterization}

\begin{figure}
\includegraphics[width=1.0\columnwidth,origin=b]{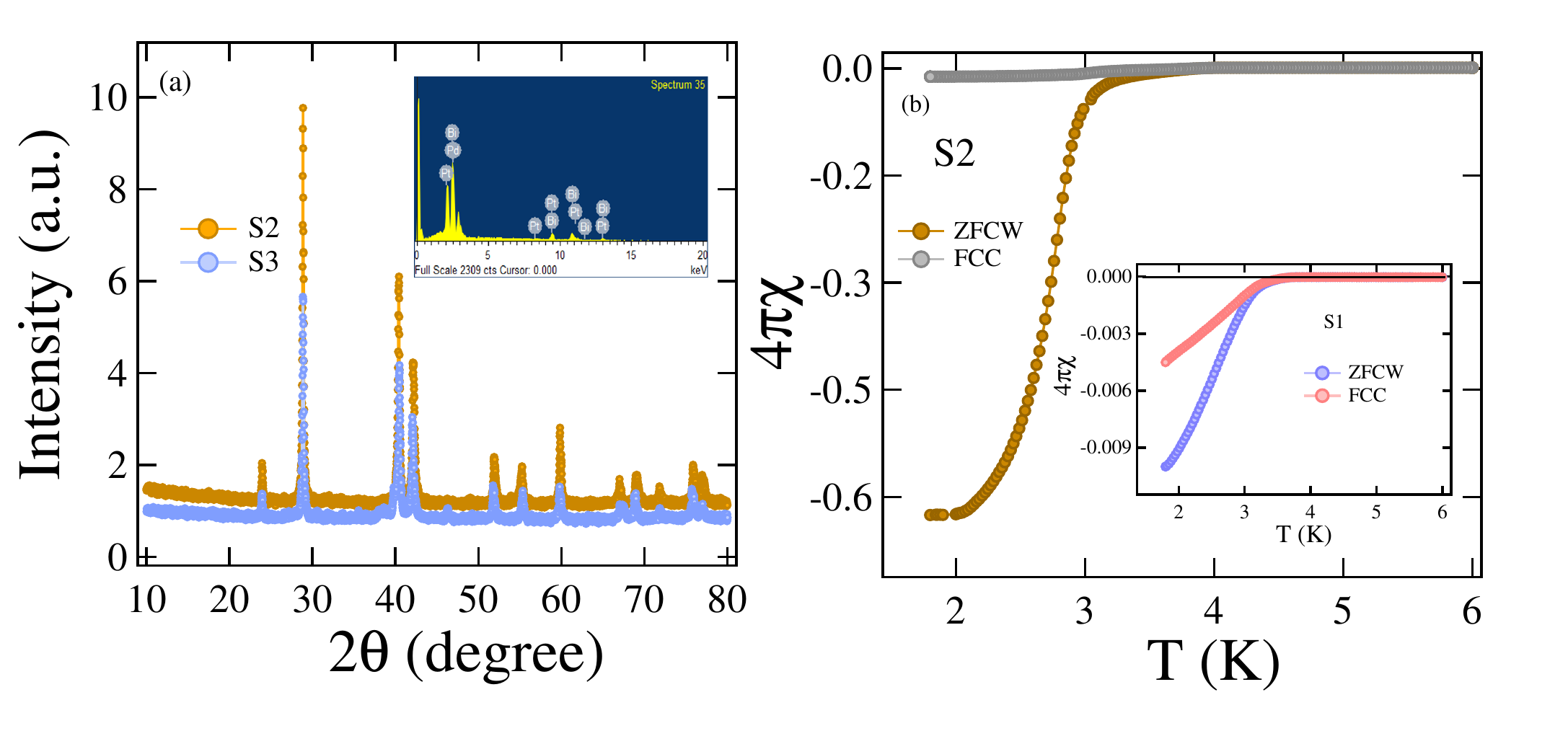}
\caption{\label{Fig9:char} (a) Shows the XRD comparison of powder sample having two different cooling rate 0.5$\degree$C/hr (S2) and 0.4$\degree$C/hr (S3) while preparation, with the inset shows the EDS pattern of S3 sample. (b) Shows the magnetic measurement compilation of S2 and S1 in the inset, whereas S3 is shown in \figref{Fig4:MT}(a).}
\end{figure}

Superconducting volume fraction in Bi$_2$PdPt is highly dependent on the sample cooling procedure during preparation. Three different samples made, such as S1, cooled at 1.0$\degree$C/hr, S2 cooled at 0.5$\degree$C/hr, and S3 cooled at 0.4$\degree$C/hr while preparing the sample. XRD of the two samples S2 and S3 is shown in \figref{Fig9:char}(a). The composition of one of the samples extracted from energy dispersive spectroscopy (EDS) is also shown in the inset of \figref{Fig9:char} (a), clearly indicating the Pt presence. The results from the magnetic characterization of S1 and S2 are presented in \figref{Fig9:char}(b). The observed superconducting transition temperature is the same for all three samples; however superconducting shielding fraction 96$\%$ is the maximum for S3. S2 only shows 60$\%$ volume fraction with less than 10$\%$ fraction is observed in S1, implying samples cooled only at a specific rate show a large volume fraction with the same diamagnetic onset temperature in magnetization measurement. Thus, reflecting the effect on grain boundaries and point defects of samples due to different cooling procedures \cite{PST}.

\subsection{Normal state properties}

\begin{figure}
\includegraphics[width=1.0\columnwidth]{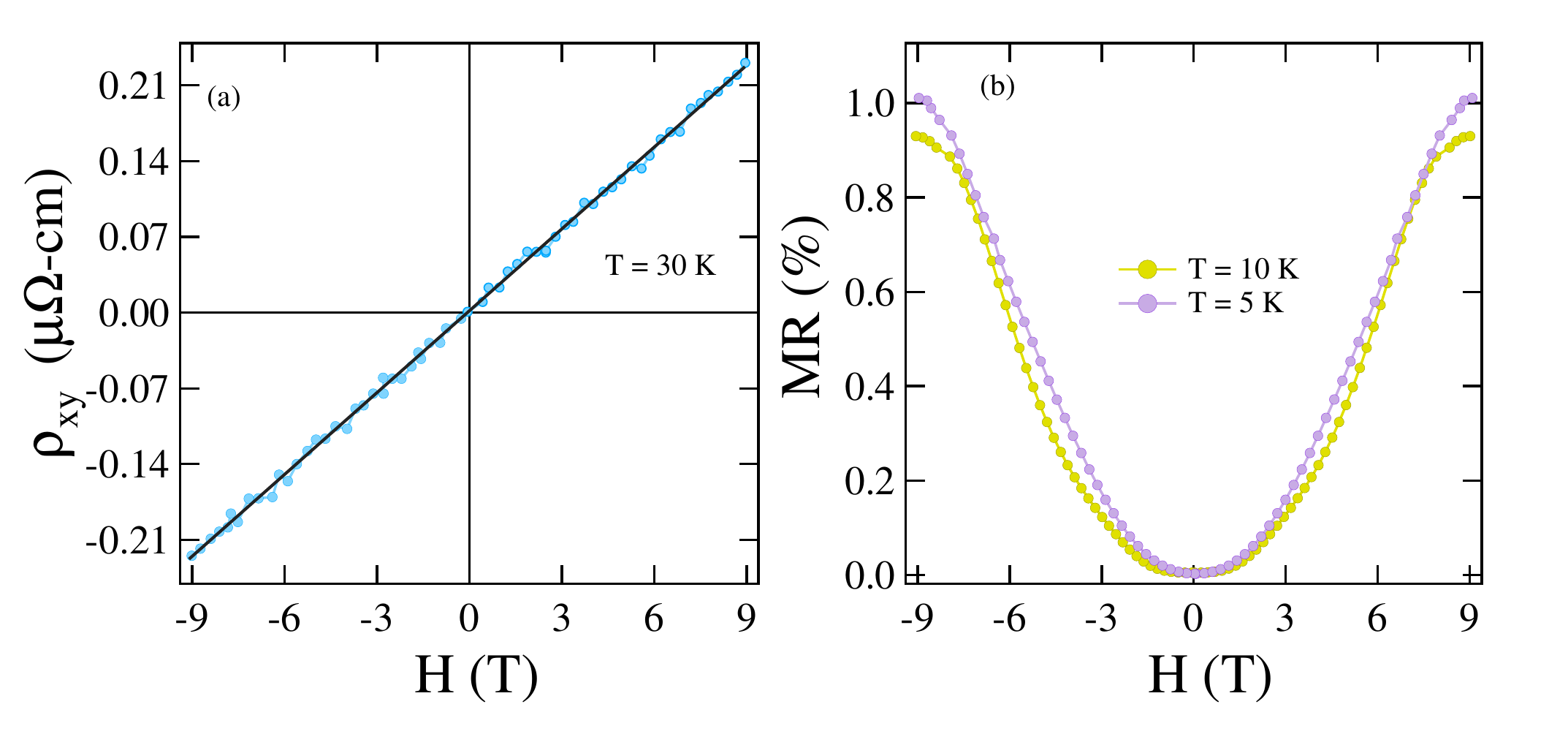}
\caption{\label{Fig10:NR} (a) Hall resistivity at $T$ = 30 K under field variation of $\pm$ 9 T. (b) Represent the MR data taken at $T$ = 5 K and $T$ = 10 K on Bi$_2$PdPt crystal.} 
\end{figure}

To perceive the normal state properties of Bi$_2$PdPt, magnetoresistance (MR) and normal hall effect were also measured. The hall measurements were done on Bi$_2$PdPt under varying magnetic field of $\pm$ 9 T at 30 K to extract the carrier density as shown in the left panel of \figref{Fig10:NR}(a), where a linear variation with magnetic field represents the small contribution from the longitudinal resistivity. The hall coefficient, $R_H$, was determined by the slope of the variation $\rho_{xy}$ with both +$H$ and $-H$ fields. The estimated $R_H$ = 2.52(2) $\times$ 10 $^{-4}$ cm$^3$/C, and the positive sign indicate the hole as the dominant charge carrier in the sample. By simply using the single band expression $n = 1/R_{H}q$, the electron concentrations was calculated, $n$ = 2.45(2) $\times$ 10$^{28} \textit{m}^{-3}$. The carrier density order matches with the other reported for $\alpha$-PdBi and PdBi$_2$ \cite{n}. MR was measured in the longitudinal configuration under a varying magnetic field of 9 T at different temperatures. MR with magnetic field dependence is defined as $\frac{\rho_{xx}(H)-\rho_{xx}(0)}{\rho_{xx}(0)}$. A significant small MR value is observed in our sample \figref{Fig10:NR}(b).


\begin{thebibliography}{References}

\bibitem{Ts} A. P. Schnyder, S. Ryu, A. Furusaki, and A. W. W. Ludwig, Phys. Rev. B 78, 195125 (2008).

\bibitem{ts1} X. L. Qi, T. L. Hughes, S. Raghu, and S.C. Zhang, Phys. Rev. Lett. 102, 187001 (2009).

\bibitem{ts2} X. L. Qi and S.C. Zhang, Rev. Mod. Phys. 83, 1057 (2011).

\bibitem{Mf1} M. Sato, and Y. Ando, Reports on Progress in Phys. 80, 076501 (2017).

\bibitem{MF}L. Fu and C. L. Kane, Phys. Rev. Lett. 100, 096407 (2008).

\bibitem{p} A. Yu. Kitaev, Physics-uspekhi 44, 131 (2001).

\bibitem{Bi-sb} L. Fu, and C. L. Kane, Phys. Rev. B 76, 045302 (2007).

\bibitem{Bi-se} H. Zhang,  C. X. Liu, X. L. Qi, X. Dai, Z.Fang, and  S. C. Zhang, Nature Physics, 5, 438 (2009).

\bibitem{TBS} T. Sato, K. Segawa, H. Guo, K. Sugawara, S. Souma, T. Takahashi, and Y. Ando, Phys. Rev. Lett. 105, 136802 (2010).

\bibitem{TBS1} K. Kuroda, M. Ye, A. Kimura, S. V. Eremeev, E. E. Krasovskii, E. V. Chulkov, Y. Ueda, K. Miyamoto, T. Okuda, K. Shimada, H. Namatame, and M. Taniguchi, Phys. Rev. Lett. 105, 146801 (2010).

\bibitem{TBT} Y. L. Chen, Z. K. Liu, J. G. Analytis, J.-H. Chu, H. J. Zhang, B. H. Yan, S.-K. Mo, R. G. Moore, D. H. Lu, I. R. Fisher, S. C. Zhang, Z. Hussain, and Z.-X. Shen, Phys. Rev. Lett. 105, 266401 (2010).

\bibitem{BPST} S. Souma, K. Eto, M. Nomura, K. Nakayama, T. Sato, T. Takahashi, K. Segawa, and Y. Ando, Phys. Rev. Lett. 108, 116801 (2012).

\bibitem{PBT} K. Kuroda, H. Miyahara, M. Ye, S. V. Eremeev, Yu. M. Koroteev, E. E. Krasovskii, E. V. Chulkov, S. Hiramoto, C. Moriyoshi, Y. Kuroiwa, K. Miyamoto, T. Okuda, M. Arita, K. Shimada, H. Namatame, M. Taniguchi, Y. Ueda, and A. Kimura, Phys. Rev. Lett. 108, 206803 (2012).

\bibitem {HH} Y. Nakajima, R. Hu, K. Kirshenbaum, A. Hughes, P. Syers, X. Wang, and J. Paglione, Science advances 1, e1500242 (2015).

\bibitem{a-PB2} Y. Zhou, X. Chen, C. An, Y. Zhou, L. Ling, J. Yang, C. Chen, L. Zhang, M. Tian, Z. Zhang, and Z. Yang, Phys. Rev. B 99, 054501 (2019).

\bibitem{a-PB2_1} S. Mitra, K. Okawa, S. K. Sudheesh, T. Sasagawa, Jian-Xin Zhu, and Elbert E. M. Chia, Phys. Rev. B 95, 134519 (2017).

\bibitem{a-PB2_2} K. Dimitri, M. M. Hosen, G. Dhakal, H. Choi, F. Kabir, C. Sims, D. Kaczorowski, T. Durakiewicz, J.X. Zhu, and M. Neupane, Phys. Rev. B 97, 144514 (2018).

\bibitem{a-PB2_3} H. Choi, M. Neupane, T. Sasagawa, Elbert E. M. Chia, and J. X. Zhu, Phys. Rev. M 1, 034201 (2017).

\bibitem{BP1} Z. Sun, M. Enayat, A. Maldonado, C. Lithgow, E. Yelland, D. Peets, D. C. Peets, A. Yaresko, A. P. Schnyder, and P. Wahl, Nat. Commun. 6, 1 (2015).

\bibitem{BP2} A, Yaresko, A. P. Schnyder, H. M. Benia, C.M. Yim, G. Levy, A. Damascelli, C. R. Ast, D. C. Peets, and P. Wahl, Phys. Rev. B 97, 075108 (2018).

\bibitem{b-PB2_A} M. Sakano, K. Okawa, M. Kanou, H. Sanjo, T. Okuda, T. Sasagawa and K Ishizaka, Nat. Commun. 6, 8595 (2015).

\bibitem{b-PB2_s} T. Xu, B. T. Wang, M. Wang, Q. Jiang, X. P. Shen, B. Gao, M. Ye, and S. Qiao, Phys. Rev. B 100, 161109(R) (2019).

\bibitem{b_PB2} J. Kacmarcik, Z. Pribulova, T. Samuely, P. Szabo, V. Cambel, J. Soltys, E. Herrera, H. Suderow, A. Correa-Orellana, D. Prabhakaran, and P. Samuely, Phys. Rev. B, 93, 144502 (2016).

\bibitem{Msc} P. K. Biswas, D. G. Mazzone, R. Sibille, E. Pomjakushina, K. Conder, H. Luetkens, C. Baines, J. L. Gavilano, M. Kenzelmann, A. Amato, and E. Morenzoni, Phys. Rev. B 93, 220504(R) (2016).

\bibitem{Msc1} E. Herrera, I. Guillamon, J. A. Galvis, A. Correa, A. Fente, R. F. Luccas, F. J. Mompean, M. Garcia-Hernandez, S. Vieira, J. P. Brison, and H. Suderow, Phys. Rev. B 92, 054507 (2015).

\bibitem{Ptbi} C. Q. Xu, X. Z. Xing, Xiaofeng Xu, Bin Li, B. Chen, L. Q. Che, Xin Lu, Jianhui Dai, and Z. X. Shi, Phys. Rev. B 94, 165119 (2016).

\bibitem{Ptbi2}J. Wang, X. Chen, Y. Zhou, C. An, Y. Zhou, C. Gu, M. Tian, and Z. Yang, Phys. Rev. B 103, 014507 (2021).

\bibitem{TI} M.Z. Hasan, and C. L. Kane, Rev. Mod. Phys. 82(4), 3045 (2010).

\bibitem{socsc} D.F. Shao, X. Luo, W. J. Lu, L. Hu, X. D. Zhu, W. H. Song, X. B. Zhu, and Y. P. Sun. Scientific Reports 6, 1 (2016).

\bibitem{fp} J. Rodríguez-Carvajal, Phys. B: Cond. Matt. 192, 55 (1993).

\bibitem {WT} A. H. Wilson, Theory of Metals (Cambridge University Press, Cambridge, England, 1958).

\bibitem{SCbp} B. Joshi, A. Thamizhavel, and S. Ramakrishnan, Phys. Rev. B 84, 064518 (2011).

\bibitem{Wm} H. Wiesmann, M. Gurvitch, H. Lutz, A. K. Ghosh, B. Schwarz, M. Strongin, P. B. Allen, and J.W. Halley, Phys. Rev. Lett. 38, 782 (1977).


\bibitem{bg} G. Grimvall, The Electron-Phonon Interaction in Metals (North-Holland, Amsterdam, 1981).

\bibitem{scpb2} Y. Imai, F. Nabeshima, T. Yoshinaka, K. Miyatani, R. Kondo, S., Komiya, I. Tsukada, and A. Maeda, Jour. of the Phys. Soc. of Jap. 81, 113708 (2012).

\bibitem{p-pb2_1}G. Pristas, Mat. Orendac, S. Gabani, J. Kacmarcik, E. Gazo, Z. Pribulova, A. Correa-Orellana, E. Herrera, H. Suderow, and P. Samuely, Phys. Rev. B 97, 134505 (2018).

\bibitem{mcm} W. L. McMillan, Phys. Rev. 167, 331 (1968).

\bibitem{Pauli_1} B. S. Chandrasekhar, Appl. Phys. Lett. 1, 7 (1962).
  
\bibitem{Pauli_2} A. M. Clogston, Phys. Rev. Lett. 9, 266 (1962).
  
\bibitem{WHH_1} E. Helfand, and N. R. Werthamer, Phys. Rev. 147, 288 (1966).
  
\bibitem{WHH_2} N. R. Werthamer, E. Helfand, and P. C. Hohenberg, Phys. Rev. 147, 295 (1966).
  
\bibitem{maki} K. Maki, Phys. Rev. B 148, 362 (1966).
  
\bibitem{Coh_Leng} M. Tinkham, Introduction to Superconductivity, 2nd ed. (McGraw-Hill, New York, 1996).

\bibitem{MIB} T. Klimczuk, F. Ronning, V. Sidorov, R. J. Cava, and J.D. Thompson, Phys. Rev. Lett. 99, 257004 (2007).
   
\bibitem{Umera_Ref} D. A. Mayoh, J. A. T Barker, R. P. Singh, G. Balakrishnan, D. McK. Paul, and M. R. Lees, Phys. Rev. B 96, 064521 (2017).

\bibitem{Umera} Y. J. Uemura, G. M. Luke, B. J. Sternlieb, J. H. Brewer, J. F. Carolan, W. N. Hardy, R. Kadono, J. R. Kempton, R. F. Kiefl, S. R. Kreitzman, P. Mulhern, T. M. Riseman, D. L. Williams, B. X. Yang, S. Uchida, H. Takagi, J. Gopalakrishnan, A. W. Sleight, M. A. Subramanian, C. L. Chien, M. Z. Cieplak, G. Xiao, V. Y. Lee, B. W. Statt, C. E. Stronach, W. J. Kossler, and X. H. Yu, Phys. Rev. Lett. 62, 2317 (1989).
 
\bibitem{Tf} A. D. Hillier and R. Cywinski, Appl. Magn. Reson. 13, 95 (1997).

\bibitem{Unconv_1} K. Hashimoto, K. Cho, T. Shibauchi, S. Kasahara, Y. Mizukami, R. Katsumata, Y. Tsuruhara, T. Terashima, H. Ikeda, M. A. Tanatar, H. Kitano, N. Salovich, R. W. Giannetta, P. Walmsley, A. Carrington, R. Prozorov, and Y. Matsuda, Science 336, 1554 (2012).
 
\bibitem{Unconv_2} R. Khasanov, H. Luetkens, A. Amato, H. H. Klauss, Z. A. Ren, J. Yang, W. Lu, and Z. X. Zhao, Phys. Rev. B 78, 092506 (2008).

\bibitem{pb} K. Kudo, H. Y. Nguyen, C. Oh, K. Takaki, and M. Nohara, Jour. of the Phys. Soc. of Jap. 90, 063706 (2021).

\bibitem{bi2pd} E. Herrera, I. Guillamon, J. A. Galvis, A. Correa, A. Fente, R. F. Luccas, F. J. Mompean, M. Garcia-Hernandez, S. Vieira, J. P. Brison, and H. Suderow, Phys. Rev. B 92, 054507 (2015).

\bibitem{tmc} M. Mandal, M. and R. P. Singh, R.P, Jour. of Phys. Cond. Matt. 33, 135602 (2021).

\bibitem{BiHH} P. Orest, D. Kaczorowski, and P. Wisniewski. Scientific Reports 5, 1 (2015).

\bibitem{sku} A. Slebarski, M. M. Maska, M. Fijalkowski, C. A. McElroy, and M. B. Maple, Jour. Alloys Comp. 646, 866 (2015).

\bibitem{PST} W. Liu , S. Li, H. Wu , N. Dhale, P. Koirala, and B. Lv,  Phys. Rev. M 5, 014802 (2021).

\bibitem{n} K. Zhao, B. Lv, Y. Y. Xue, X. Y. Zhu, L. Z. Deng, Z. Wu, and C. W. Chu, Phys. Rev. B 92, 174404 (2015).


\end{thebibliography}
\end{document}